\documentclass[prl, aps,amsmath,amssymb,amsfonts, superscriptaddress,twocolumn]{revtex4}
\usepackage[german,american,english]{babel}
\usepackage{graphicx}
\usepackage{graphics}
\usepackage{dcolumn}
\usepackage{multirow}
\usepackage{bm}
\usepackage{latexsym}
\usepackage{amssymb}
\usepackage{amsmath}
\usepackage{amsfonts}
\usepackage{layout}
\usepackage{verbatim}
\usepackage{epsfig}
\usepackage{graphicx}
\usepackage{amsbsy}
\usepackage{xcolor}
\usepackage[caption=false]{subfig}
\newcommand{\bea}{\begin{eqnarray*}}
	\newcommand{\eea}{\end{eqnarray*}}
\newcommand{\bne}{\begin{equation*}}
\newcommand{\ede}{\end{equation*}}

\newcommand{\bnen}{\begin{equation}}
\newcommand{\eden}{\end{equation}}
\newcommand{\bean}{\begin{eqnarray}}
\newcommand{\eean}{\end{eqnarray}}
\newcommand{\bsen}{\begin{subequations}}
	\newcommand{\esen}{\end{subequations}}

\newcommand{\bna}{\begin{array}}
	\newcommand{\eda}{\end{array}}
\newcommand{\bnm}{\begin{enumerate}}
	\newcommand{\edm}{\end{enumerate}}

\newcommand {\ket} [1] {| #1 \rangle}
\newcommand {\bkt} [1] {\langle #1 \rangle}

\newcommand {\pd} [2] {\frac{\partial #1}{\partial #2}}

\begin{document}
	
\title{Intrinsic torque on the orbital angular momentum in an electric field}
\author{Rhonald Burgos Atencia}
\affiliation{School of Physics, The University of New South Wales, Sydney 2052, Australia}
\affiliation{ARC Centre of Excellence in Low-Energy Electronics Technologies, UNSW Node, The University of New South Wales, Sydney 2052, Australia}
\author{Daniel P. Arovas}
\affiliation{Department of Physics, University of California at San Diego, La Jolla, CA 92093-0319,USA}
\author{Dimitrie Culcer}
\affiliation{School of Physics, The University of New South Wales, Sydney 2052, Australia}
\affiliation{ARC Centre of Excellence in Low-Energy Electronics Technologies, UNSW Node, The University of New South Wales, Sydney 2052, Australia}

\begin{abstract}
Orbitronics harnesses non-equilibrium densities and flows of electrons’ orbital angular momentum (OAM). Although the OAM must be long lived to be useful in information processing, the mechanisms leading to OAM non-conservation remain unknown. Here we show that an electric field induces an intrinsic torque on the OAM
mediated by the quantum metric tensor, without spin-orbit coupling or magnetic fields. The torque vanishes in pseudospin-1/2 systems with particle-hole symmetry but is nonzero in the presence of nontrivial textures. We discuss implications for topological materials and strategies for experimental observation. 
\end{abstract}
\date{\today}
\maketitle

\textit{Introduction}. The past decade has witnessed unprecedented interest in \textit{orbitronics}, whose focus is generating non-equilibrium densities and currents of electrons' orbital angular momentum (OAM). Orbitronics emphasizes light materials, ideal for mass production \cite{DingShilei2020, LeeSoogil2021, LeeDongjoon2021, DingShilei2022, ChoiYoungGwan2023, HayashiHiroki2023}, in which magnetic interactions rely on orbital degrees of freedom given that spin-orbit coupling is either weak or absent altogether. The OAM, associated with an electron's wave packet motion about its center of mass \cite{Yafet1961, Yafet-1963, ChangMingChe1996, GaneshSundaram1999, Vanderbilt2018}, affects the valley-dependent $g$-factor of graphene and MoS$_2$ structures \cite{Rostami2015, GuitierrezRubio2016, Overweg2018, AngelikaKnothe2018, YongjinLee2020, Tong2021}. In out-of-equilibrium setups the OAM is investigated for information processing and memory applications, including the gyrotropic magnetic effect \cite{ZhongShudan2016, Rou2017, ElioKonig2019}, the orbital Edelstein effect \cite{YodaTaiki2015, YodaTaiki2018, SalemiLeandro2019, TarikCysne2021, TarikCysne2023, ShinadaKoki2023}, the valley Hall effect \cite{XiaoDi2007, MakKF2014, BeconciniMichael2016, LeeJieun2016, BhowalSayantika2021}, the orbital Hall effect \cite{Bernevig2005, BhowalSayantika2020, CysneTarik2021, BhowalSayantika2021, PezoArmando2022, CysneTarik2022, kazantsev2023, GiacomoSala2023, IgorLyalin2023}, and the associated orbital torque \cite{GoDongwook2018, Dongwook2020-I, GoDongwook2020-II}, among others \cite{GaoYang2018, Lan2023}. 

Orbitronic devices require a transparent operational criterion, which does not exist at present. Whereas the equilibrium OAM is well understood in a perfect crystal \cite{ChangMingChe1996, GaneshSundaram1999, XiaoDi2005, Culcer2005, MingCheChang2008, XiaoDi2010, GaoYang2015, LiangDong2018, Aryasetiawan2019, XiaoCong2021, PalOjasvi2022, PalOjasvi2023, XiaoDi2005, Thonhauser2005, CeresoliDavide2006, ShiJunren2007, Thonhauser2011, RaffaeleResta2012, Nourafkan2014, Aryasetiawan2016}, interest in the OAM is overwhelmingly in out-of-equilibrium systems, and the extent to which information encoded in the OAM is conserved is not known. Three fundamental reasons make an understanding of OAM non-conservation an urgent imperative: (i) For information processing and memory it is vital to know how long the OAM can live; (ii) In orbitronics devices orbital dynamics is almost always entangled with spin dynamics, and charge-orbital-spin conversion occurs at interfaces; (iii) Knowledge of OAM non-conservation enables conserved orbital currents and torques to be defined and studied in a meaningful way. The extensive literature on the spin-Hall effect has shown that the transport of a non-conserved quantity is meaningless, and that the conserved spin current can be much smaller than the conventional spin current. Switching bits by means of the orbital Hall effect makes the tacit assumption that the quantity being transported is conserved, which is not guaranteed out of equilibrium. The importance of intrinsic, electric-field induced, non-conservation of the valley density, and, by extension, of the OAM density, has been recently recognized and discussed in Ref.~\cite{kazantsev2023}. The problem is highly non-trivial for the OAM, as can be seen by direct comparison with the spin. Spin-dependent terms, leading to spin non-conservation, can usually be read off immediately in the Hamiltonian, and can often be encapsulated in a momentum-dependent effective magnetic field as is customarily done for the Rashba and Dresselhaus interactions. On the other hand orbital dynamics is \textit{hidden} in the Hamiltonian, leading to the following fundamental open questions: \textit{When is the OAM at each wave vector ${\bm k}$ not conserved? When is it not conserved globally? What intrinsic mechanisms lead to OAM non-conservation and how are they related to the underlying topology?}

In this paper we address the fundamental questions above and demonstrate that the OAM is in general not conserved even in the absence of spin-orbit coupling and magnetic fields. Focussing on a generic solid in an electric field we show that OAM non-conservation stems from the topological texture inherent in the Bloch wave functions, and that the leading contribution comes from the quantum metric tensor, with an additional contribution in multi-band systems arising from the inter-band velocity. Surprisingly both contributions to the torque depend on the group velocity. This differs from the equilibrium situation as well as from the orbital Hall effect, where only the inter-band velocity enters. We study non-conservation at each ${\bm k}$, as well as the presence of a net, global torque density. We illustrate these with reference to a generic model of tilted massive Dirac fermions, where the torque is determined by the magnitude and direction of the tilt. Based on this we show that no torque on the OAM is expected in transition metal dichalcogenides even when spin-orbit coupling is accounted for. We argue that there is no torque on the OAM in bulk topological insulators, leading to the important conclusion that the conventional orbital Hall current in these materials conserved, even as the conventional spin-Hall current is not conserved. Whereas our examples focus on topological materials, we stress that our results apply to all solids. 

\textit{Torque on OAM}. We focus on the intrinsic response of the torque operator. We assume the system is spatially homogeneous and the bands are non-degenerate. The operator form of torque reads $\tau_{\gamma}=\dot{L}_{\gamma} = \epsilon_{\alpha\beta\gamma} \frac{1}{2} \{\hat{r}_\alpha,\hat{a}_\beta \}$, where the acceleration operator $\hat{a}_\beta=(i/\hbar)[H, \hat{v}_{\beta}]$, with $\hat{v}_{\beta}$ the velocity operator. The torque is therefore given by the expectation value of the acceleration moment, in the same way as in classical mechanics. With the system described by a density matrix $\rho$, the expectation value of the torque is given by $\langle \tau_{\gamma} \rangle = \epsilon_{\alpha\beta\gamma}\frac{1}{2} {\rm Tr}[ \rho\{\hat{r}_\alpha,\hat{a}_\beta \}]
=\epsilon_{\alpha\beta\gamma}\frac{1}{2} {\rm Tr}[ \{\hat{r}_\alpha,\rho \} \hat{a}_\beta  ]$, where $\rho$ is the density matrix. The crystal momentum representation is given by $\ket{\Psi_{m{\bm k}}} = e^{i{\bm k}\cdot{\bm r}} \ket{u_{m{\bm k}}}$, where $|u^{m}_{\bm k} \rangle$ is the periodic part of the Bloch wave function. The velocity and acceleration operators are diagonal in wave vector. The matrix elements of the position operator are evaluated using ${\bm r} \ket{\Psi_{m{\bm k}}} \rightarrow [-i (\partial/\partial{\bm k}) \, e^{i{\bm k}\cdot{\bm r}}]\, \ket{u_{m{\bm k}}}$. Following straightforward steps outlined in the Supplement \cite{Supplement}, we obtain
\begin{equation}\label{eq:dA}
	\arraycolsep 0.3ex
	\begin{array}{rl}
		\displaystyle \bkt{\tau_{\gamma}} = & \displaystyle \epsilon_{\alpha\beta\gamma} {\rm tr} \int \frac{d^dk}{(2\pi)^{d}} \, a^{\bm k}_{\beta} \Xi^{{\bm k}, \alpha},
	\end{array}
\end{equation}
where tr is the sum over band indices $m,n$, and 
\begin{equation}\label{eq:xi}
	\arraycolsep 0.3ex
	\begin{array}{rl}
		\displaystyle \Xi^{{\bm k}, \alpha}_{nm} = &\displaystyle \frac{i}{2} \, \bigg[ \bigg(\pd{\tilde{\rho}^{nm}_{{\bm k} + \frac{\bm Q}{2}, {\bm k} - \frac{\bm Q}{2}}}{Q_\alpha}\bigg) - \bigg(\pd{\tilde{\rho}^{nm}_{{\bm k} - \frac{\bm Q}{2}; {\bm k} + \frac{\bm Q}{2}}}{Q_\alpha}\bigg) \bigg]_{{\bm Q} \rightarrow 0} \\ [3ex]
  \displaystyle + & \displaystyle\frac{1}{2} \, \{\mathcal{R}_\alpha, \tilde{\rho}_{\bm k} \}_{nm}.
	\end{array}
\end{equation}
The Berry connection 
$\mathcal{R}^{mm'}_{\bm k, \alpha}=i\langle u^{m}_{\bm k} | \partial u^{m'}_{\bm k}/\partial k_{\alpha} \rangle $ appears due to the wave vector dependence of the basis functions. Using this method we immediately recover the correct semiclassical expressions for the OAM ~\cite{GaneshSundaram1999, MingCheChang2008}, and we 
find that the torque is zero in equilibrium.

We recall that the equilibrium OAM involves only the band off-diagonal elements of the position operator \cite{ChangMingChe1996, GaneshSundaram1999, XiaoDi2005, Culcer2005, MingCheChang2008, XiaoDi2010}. Obtaining the correct torque on the OAM also requires the band-diagonal elements of ${\bm r}$, which lead to group velocity, Fermi surface, and dipolar effects. The diagonal matrix elements of ${\bm r}$ have been extremely challenging for Bloch electrons \cite{RestaRaffaele1998, RaffaeleResta2012, Resta2018, Vanderbilt2018} -- they are straightforward to treat when appearing in a commutator, but make the calculation highly non-trivial inside an anti-commutator, as is the case for the torque. These matrix elements are captured by Eq.~\ref{eq:xi} above, which we believe represents a substantial technical advance.

\textit{Linear response}. The density matrix satisfies the quantum Liouville equation $\partial_t \rho = (i/\hbar)[H, \rho]$, where $H$ is the Hamiltonian. In a constant, uniform electric field $H = H_0 + H_{E}$, where $H_0$ is the band Hamiltonian and $H_E = e {\bm E} \cdot \bm r$ is the electrostatic potential. We are interested in the expectation value of the torque in an electric field, that is, Tr ${\bm \tau} \rho$ to first order in ${\bm E}$. This expectation value has two parts, since both the torque operator and the density matrix have contributions to zeroth and first orders in the electric field respectively. The torque operator has a contribution from $H_0$, which we shall denote by ${\bm \tau}_0$, and a contribution from $H_E$, denoted by ${\bm \tau}_E$. Similarly, $\rho = \rho_0 + \rho_E$, where $\rho_0$ is the equilibrium density matrix and $\rho_E$ is the correction to linear order in the electric field, found below. The equilibrium density matrix is diagonal in the band index, its matrix elements being the Fermi-Dirac distribution $f^{m}_{\bm k}$ for each band. In equilibrium it is easy to check that Tr ${\bm \tau}_0 \rho_0 = 0$, as expected. The two contributions we seek to determine are Tr ${\bm \tau}_E \rho_0$ and Tr ${\bm \tau}_0 \rho_E$. In the first contribution, ${\bm \tau}_E$ is evaluated straightforwardly from the commutator $(i/\hbar) \, [e {\bm E} \cdot \bm r, {\bm v}]$, and the result is weighted by the equilibrium density matrix. The strategy for evaluating the second contribution, Tr ${\bm \tau}_0 \rho_E$, can be summarized as follows: (i) Determine $\rho_E$ from the quantum Liouville equation; (ii) In linear response $\rho_E$ is expressed in terms of the equilibrium density matrix $\rho_0$; (iii) Hence the trace Tr ${\bm \tau}_0 \rho_E$ can be re-expressed in terms of $\rho_0$; (iv) The quantity multiplying $\rho_0$ is the second contribution to the torque.

To find $\rho_E$ we resort to the quantum Liouville equation. In linear response the band- and ${\bm k}$-off-diagonal intrinsic distribution $\rho^{mm'}_{E; {\bm k}{\bm k}'}$ is found through an immediate generalization of Refs.~\cite{BurgosPRR2022, Culcer2017, Sekine2017}:
\begin{align}
\label{Eq:PurelyIntrinsicOffDiagonalLinearDistribution}
\rho^{nn'}_{E{\bm k} {\bm k}'}
&=\frac{eE_{b} \, (f^{n}_{\bm k} - f^{n'}_{{\bm k}'}) \, \mathcal{R}^{nn'}_{{\bm k} {\bm k}'; b}}{\epsilon^{n}_{{\bm k}} -\epsilon^{n'}_{{\bm k}'}},
\end{align}
where $\mathcal{R}^{nn'}_{{\bm k} {\bm k}';b}$ is now the matrix element of the Berry connection off-diagonal in wave vector. Using this result, we can express the torque induced by an electric field at each wave vector ${\bm k}$ in the form $\displaystyle \sum_{n{\bm k}} \tau_{nn; {\bm k}}^i f^{n}_{\bm k}$, where
\begin{widetext}
\begin{align}
\label{Eq:generatorque}
\tau^i_{nn;{\bm k}}
&=
\displaystyle \frac{e E_a}{\hbar }\epsilon_{ijk} \,\bigg( \sum_{m \ne n} 
{\rm Im}[\mathcal{Q}^{nm}_{ka} ] \, (v^j_{mm} + v^j_{nn}) - \hbar\sum_{m \ne n \ne l}  
{\rm Re}\bigg[
\frac{\mathcal{R}^{a}_{nl} \left(v^k_{lm} v^j_{mn} + v^j_{lm} v^k_{mn} \right)}{\varepsilon_n - \varepsilon_l}
\bigg] \bigg)
\end{align}
\end{widetext}
where $\mathcal{Q}^{nn'}_{jb}=\mathcal{R}^{nn'}_{\bm k,j}  \mathcal{R}^{n'n}_{\bm k,b}$ is the quantum metric tensor, Im and Re denote the imaginary and real parts respectively, and $\bm{v}^{nn}_{\bm k}$ is the diagonal (group) velocity.  

\textit{Discussion}. This result, as given in Eq.~\ref{Eq:generatorque}, applies to all solids. It shows that, surprisingly, the OAM in general experiences an intrinsic torque in the absence of magnetic fields. As the models below show, it is also nonzero in the absence of spin-orbit interactions. By inspection it is immediately obvious that the second term in Eq.~\ref{Eq:generatorque} vanishes in a two-band model given the requirement that the indices $l$, $m$, $n$ be different. In such a two-band model the remaining term in Eq.~\ref{Eq:generatorque} vanishes if the model has particle-hole symmetry, implying that ${\bm L}$ is conserved.

The following important conclusions can be drawn from Eq.~\ref{Eq:generatorque}. The torque stems from: the quantum metric tensor due to the non-trivial topological texture inherent in the Bloch wave functions; the Berry connection terms due to interband coherence induced by the electric field; the group velocity, which comes from the band-diagonal terms in the position operator. The geometric nature of the torque is apparent, given that the presence of the Berry connection and quantum geometric tensor. The quantum geometric tensor can be written as $\mathcal{Q}^{mm'}_{ij}=\langle \partial_{k_i}u^{m}_{\bm k} | u^{m'}_{\bm k} \rangle \langle  u^{m'}_{\bm k} | \partial_{k_j}u^{m}_{\bm k} \rangle $. Its real part is related to the quantum metric while its imaginary part is related to the Berry curvature. Importantly, the torque involves the group velocity, which is traced to the band-diagonal matrix elements of the position operator. This is the first time that a band-diagonal quantity is shown to contribute to the OAM dynamics of Bloch electrons -- it is entirely different from the equilibrium case as well as from the orbital Hall effect, the OAM matrix elements involved are traced to the inter-band position/velocity matrix elements appearing through the Berry connection. The presence of two group velocities is understood as follows. The equilibrium intrinsic OAM is the result of inter-band matrix elements in both the velocity and the position operators. Out of equilibrium the intrinsic OAM remains an inter-band coherence phenomenon involving the coupled dynamics of pairs of bands. In fact, the full calculation requires the band off-diagonal elements of the orbital moment \cite{PezoArmando2022, CysneTarik2022, CysneTarik2021, BhowalSayantika2020, BhowalSayantika2021, XiaoDi2007}. The torque is likewise an inter-band coherence effect -- a form of Zitterbewegung -- since the acceleration also couples neighbouring bands. Hence the group velocity of each band in the pair must contribute to the torque. Physically, if the OAM is thought of as an interband coherence phenomenon \cite{Culcer2017}, 
the torque can be understood as result of the electric field mixing the OAMs from different bands.

In the general 3D case it is natural to expect a component of ${\bm \tau} \parallel {\bm L}$ and a component of $\perp {\bm L}$. Given the complexity of the resulting expressions, these would have to be evaluated for specific models. However, considerable insight can be gained by focusing on the much more straightforward 2D case, where necessarily ${\bm L} \parallel \hat{\bm z}$ and ${\bm \tau} \parallel \hat{\bm z}$. Unlike the spin, the OAM in a 2D system can only point out of the plane, and there can be no precession. Hence in 2D the OAM can only change as a result of (i) a torque parallel or antiparallel to it, as described here for the case of an electric field; or (ii) flipping as a result of random processes such as scattering. This is in sharp contrast to spin dynamics in a 2D system, in which all three components of the spin are present and the spin precesses under the action of spin-orbit coupling of e.g. the Rashba or Dresselhaus forms.

Whereas the torque at the Fermi energy is indicative of the time scales associated with the time change of the OAM, the net torque density is relevant for defining a correct OAM current, following arguments similar to Refs.~\cite{Dimi2004,Sinova2004,ShiJunren2006} for the spin current. The conventional definition of the OAM current used in the orbital Hall effect is simply the product of the OAM and velocity operators. In systems in which the angular momentum is not conserved the current needs to be augmented by a torque dipole. Our work shows that the conventional definition of the OAM current is appropriate for massless Dirac fermions as long as there is no tilt. From a symmetry point of view, the presence of a net torque density requires gyrotropic symmetry in the same way as the orbital and spin Edelstein effects 
\cite{EDELSTEIN1990,YodaTaiki2015,Borge2015,YodaTaiki2018,JohanssonAnnika2021}, given that the torque ${\bm \tau}$ has the same symmetry requirements as the angular momentum. Yet, as Eq.~\eqref{Eq:generatorque} shows, gyrotropic symmetry is not enough: the torque vanishes in a two-band system if particle-hole symmetry is present. 

A recent work \cite{kazantsev2023} has pointed out subtle non-conservation issues related to the number density in a single valley in an electric field, which in turn will affect OAM conservation. In this context we note that our calculation does not address valleys explicitly. Whereas Eq.~\ref{Eq:generatorque} is general, to obtain  physical insight one needs to apply it to a specific model, and ${\bm k} \cdot {\bm p}$ models focus on individual valleys. Hence, if one considers a single-valley ${\bm k} \cdot {\bm p}$ model, even if the model has particle-hole symmetry implying that Eq.~\ref{Eq:generatorque} vanishes, the number density in a single valley is not conserved, as pointed out in Ref.~\cite{kazantsev2023}. The interplay of OAM and valley conservation will require further investigation.

We stress that the quantity calculated here is the intrinsic torque on the conduction electrons' OAM in the presence of an electric field: it is \textit{not} related to the orbital torque studied in ferromagnetic materials \cite{GoDongwook2020I,GoDongwook2020II,BoseArnab2023,GoDongwook2023}. 
What is more, our present calculation focuses on the OAM rather than the orbital magnetic moment of Bloch electrons. The latter is known to contain an additional contribution due to the Berry curvature correction to the density of states \cite{XiaoDi2005, Vanderbilt2018}, which is not our concern in this work. However, within the density matrix formalism is possible to include such a correction as explained in Ref.\cite{Sekine2017}.


\begin{figure}[tbp]
\centering
\includegraphics[width=0.43\textwidth]{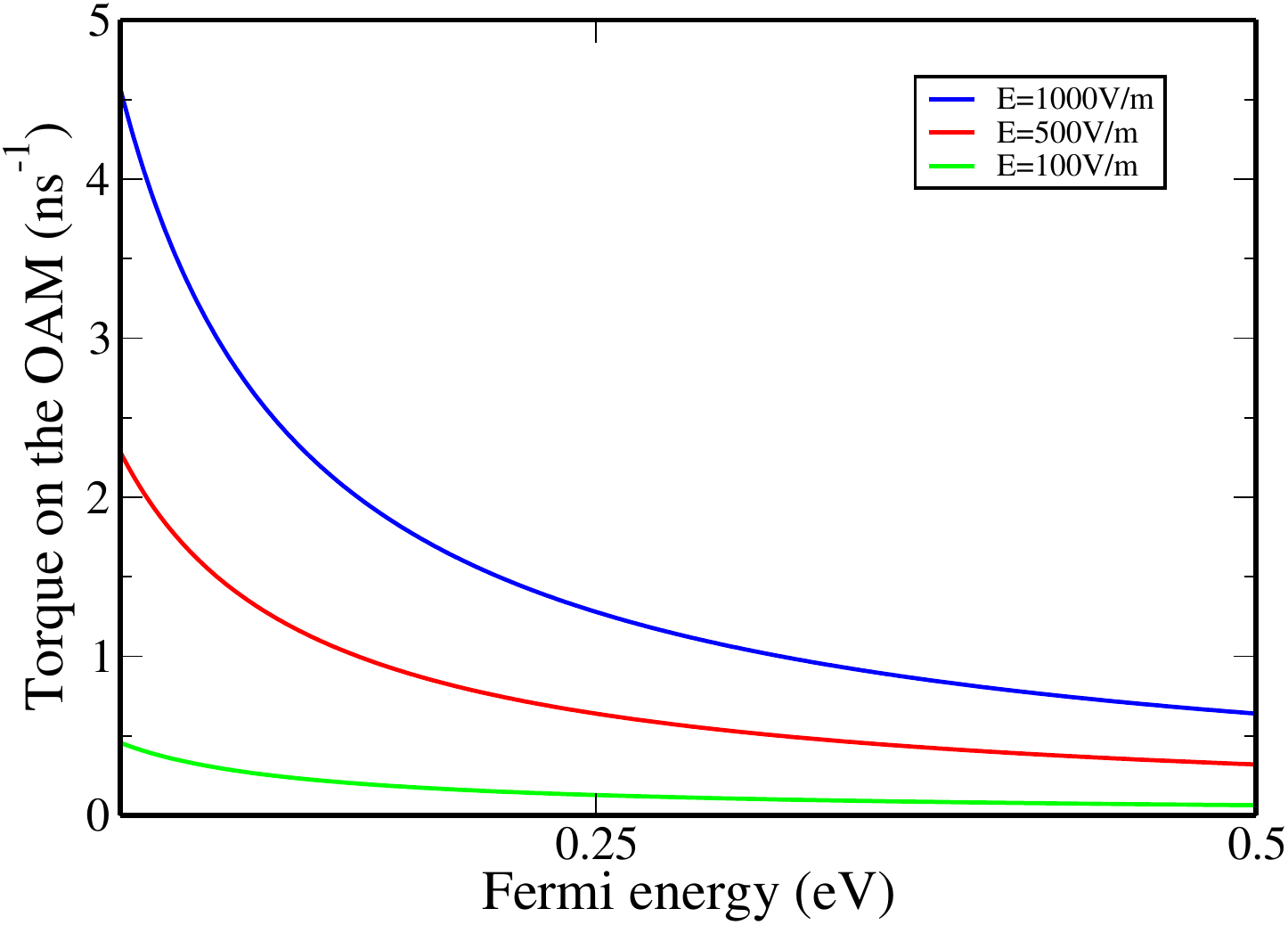}
\caption{Torque on the OAM at the Fermi energy with tilt $t=0.4$ and $v_F=1.6\times 10^{6}$m/s. Here we have plotted $(1/L) (dL/dt)$ with $1/L$ evaluated in equilibrium. In the text, for simplicity, we will call it $\omega_0(\epsilon_F)$. We used three different electric fields and a gap $\Delta=70$meV.}
\label{Fig:frequency}
\end{figure}

\textit{Applications to specific models}. We have checked that the OAM rate of change is zero for bulk topological insulators (Bi$_2$Se$_3$, Bi$_2$Te$_3$, Sb$_2$Te$_3$) described by the model given in  Ref.~\cite{LiuChaoXingPRB2010, JamesCullen2023}. In this model even the multi-band term, that is, the second term in Eq.~\ref{Eq:generatorque} is zero in view of the fact that both the conduction and the valence bands are two-fold degenerate. This leads to the important conclusion that the conventional definition of the orbital Hall current in bulk topological insulators is conserved. This is in sharp contrast to the conventional spin current, which is not conserved due to strong spin-orbit coupling. 

We have also found the OAM rate of change to be zero for transition metal dichalcogenide mono-layers (e.g. MoS$_2$) in the model of Ref.~\cite{XiaoDiPRB2012}. For this system, even though particle-hole symmetry superficially appears to be absent, the model can be broken down into two copies, each having the form of a constant term plus a $2 \times 2$ particle-hole symmetric term. Hence it is straightforward to show that Eq.~\ref{Eq:generatorque} leads to a zero OAM rate of change.

We consider next a tilted Dirac cone 
\begin{equation}
\label{Eq:model}
H_0=\hbar v_t k_x\sigma_0+\hbar v_{0x}k_x \sigma_x + \hbar v_{0y}k_y \sigma_y + \Delta \sigma_z.
\end{equation} 
The energy spectrum is  $\epsilon^{\pm}_{\bm k}=\hbar t k_x + \epsilon^{\pm}_{0k}$, where $\epsilon^{\pm}_{0k}= \sqrt{\hbar^2k^2+\Delta^2}$. We changed variable as $k_x \rightarrow v_{0x}k_x$ and $k_y \rightarrow v_{0y}k_y$ and defined the parameter $t=v_t/v_{0x}$. Although our system is $\mathcal{P}\mathcal{T}$-broken, it has the reduced symmetry $\mathcal{M}_{x}\mathcal{T}$ where $\mathcal{M}_{x}$ stands for mirror symmetry in the $x$-direction. Such a symmetry implies that only the coefficient $\kappa^{(0)}_{zx}$ is allowed. 


We find the torque to be proportional to the tilt parameter. In the context of interband coherence, we can understand this as follows: If there is no tilt the OAMs from different bands do not mix. The tilt  mixes the OAMs from the two bands and this mixing is asymmetric, leading to a net change of the OAM from the two bands. Once the torque is found, we divide this expression by the equilibrium OAM we obtain a formula for the time scale characterizing the OAM rate of change under the action of an electric field. This time scale is plotted in Fig.\eqref{Fig:frequency} as a function of the Fermi energy for several values of the applied electric field. The total torque density is found by integrating over wave vector:
\begin{equation}
\label{Eq:OrbitalTorque}
\langle \tau_{z} \rangle_E = -\frac{v_{0x}}{e}\hat{\bm t} \times \hat{\bm j}_{Hall}.
\end{equation}
Here the intrinsic anomalous Hall current \cite{BurgosPRR2022} $j^{Hall}_{y}=e^2E_x\left( \frac{\Delta}{4\pi\hbar \epsilon_{F}} \right)$ and the tilt $\hat{\bm t} =t \hat{\bm e }_{x}$ is in the $x$-direction. Explicitly the torque density takes the form $\langle \tau_{z} \rangle_E = -t e v_{0x} E_x\frac{\Delta}{4\pi\hbar \epsilon_{F}}$. This intrinsic torque is a Fermi sea effect. The fact that the torque is zero for $\mathcal{P}\mathcal{T}$-symmetric massive Dirac fermions, in which the tilt is absent, is again evident from Eq.~\ref{Eq:OrbitalTorque} since the intrinsic anomalous Hall response is only allowed in $\mathcal{P}\mathcal{T}$-broken systems.

\begin{figure}[tbp]
\centering
\includegraphics[width=0.43\textwidth]{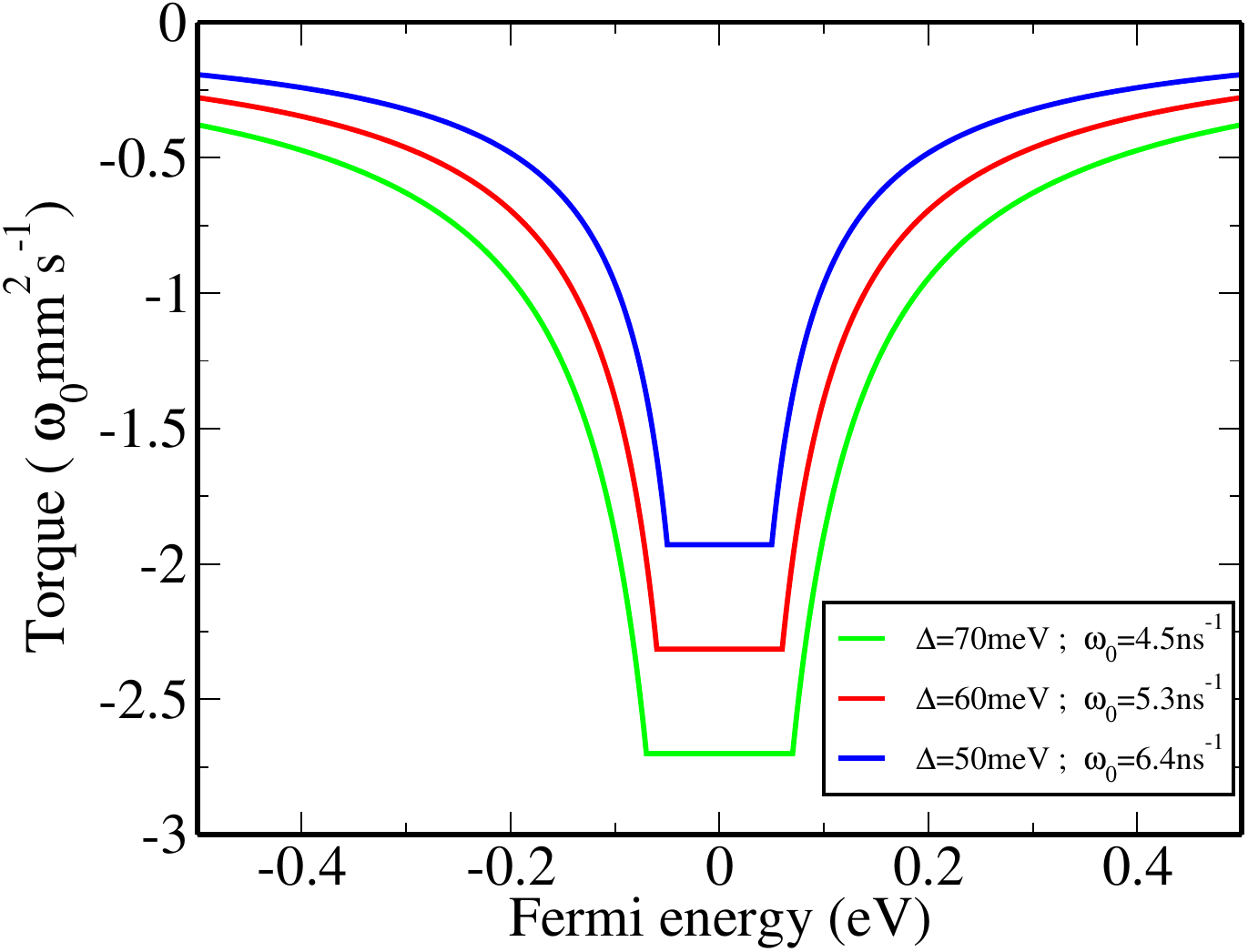}
\caption{Torque for massive Dirac fermions for different gap parameters with tilt $t=0.4$ and $v_F=1.6\times 10^{6}$m/s. The electric field $E=1000$V/m and the torque is plotted in units of $\omega_0 mm^2/s$, where $\omega_0= tev_FE /2\Delta $ is a characteristic frequency (at the bottom of the conduction band). The torque is multiplied by the approximate area of a unit cell $A \approx 1nm^2$.}
\label{Fig:torque}
\end{figure}

Since the tilt and the anomalous Hall current have opposite signs in the two valleys their contributions are additive and the torque density is nonzero in the system. The torque density is plotted in Fig.\eqref{Fig:torque} as a function of the Fermi energy. One possibility to test for the effect of the torque experimentally would be to measure the change in the valley-dependent $g$-factor of e.g. bilayer graphene or TMDs as an electric field is turned on.
We note that disorder effects are expected to be important based on recent results about side-jump and skew scattering in the orbital Hall effect \cite{HongPRL2024,tang2024}. 
Such effects will be addressed in a future publication.

\textit{Conclusions and Outlook}. We have shown that the OAM of Bloch electrons experiences a torque in an electric field. This torque does not involve magnetic fields or spin-orbit interactions, but is an inter-band coherence effect involving the quantum metric tensor as well as the group velocity, whose presence is a consequence of the band-diagonal matrix elements of the position operator. The torque vanishes in two-band systems with particle-hole symmetry but is nonzero in general. These results will affect a variety of novel materials. For example, the discovery of the anomalous Hall effect in twisted bi-layer graphene and Moir\'e systems hints at  magnetic properties associated with orbital degrees of freedom 
\cite{JianpengLiu2019, SharpeAaron2019, SerlinM2020, HeWenYu2020, JianpengLiu2021, TschirhartCL2021, ChunliHuang2021, XiaoboLu2019, SameerGrover2022}.  Although in many materials orbital and spin magnetism are expected to co-exist, orbital magnetism may dominate under certain circumstances \cite{JoDaegeun2018, KimJunyeon2021, JohanssonAnnika2021}. Our results are presented in a form valid for computational approaches based on tight-binding, density functional theory and related methods \cite{CostaMarcio2023, SalemiLeandro2022, MarzariNicola2012, LopezMG2012, CeresoliDavide2010} and also can be easily extended to multi band systems. 


\textit{Acknowledgments}. This paper has been stimulated by a series of fundamental discussions with David DiVincenzo, Roberto Raimondi, Thierry Valet, Giovanni Vignale, Eli Zeldov, Shuichi Murakami, Hiroshi Kohno, Binghai Yan, and Kamal Das. This project is supported by the Australian Research Council Centre of Excellence in Future Low-Energy Electronics Technologies (project number CE170100039).



\end{document}